\documentclass[12pt]{article}

\usepackage[margin=1in]{geometry}
\usepackage{amsmath,amssymb,amsthm}
\usepackage{graphicx}
\usepackage{hyperref}
\usepackage[numbers,sort&compress]{natbib}
\usepackage{lineno}
\usepackage{xcolor}
\usepackage[normalem]{ulem}

\begin{document}

\title{\textbf{Restricting One-Loop radiative effects in quantum gravity:
Demonstrating 4D GR as an EFT and its consistent unification with the Standard Model}}
\author{
    Farrukh A.~Chishtie\thanks{
        $^1$Peaceful Society, Science and Innovation Foundation, 204-2295 West Broadway, Vancouver, BC V6K 2E4, Canada.\\
        $^2$Department of Occupational Science and Occupational Therapy, Faculty of Medicine,  University of British Columbia, 2211 Wesbrook Mall T325, Vancouver, BC V6T 2B5, Canada.
        \texttt{email: fachisht@uwo.ca}
    }
}

\date{}
\maketitle

\begin{abstract}
In ``On restricting to one-loop order the radiative effects in quantum gravity" (Brandt, Frenkel, and McKeon, 2020) \cite{Brandt2020}, a Lagrange multiplier (LM) field is introduced into the Einstein-Hilbert action, removing all multi-loop graviton diagrams and confining quantum-gravity corrections to just one loop. The resulting one-loop effective action carries a term proportional to {$\ln(\mu/\Lambda)$}, which they suggest could be experimentally determined, hinting at direct measurements of quantum-gravity effects. We {demonstrate}, however, that {$\mu$ and $\Lambda$} emerge from a chosen renormalization scheme, not from physical observables, implying that {$\ln(\mu/\Lambda)$} signals a finite UV cutoff in this ``LM renormalization scheme.'' Although Newton's constant remains fixed (no running of {$G_N$}), the resulting logarithmic dependence encodes a limited domain of validity for General Relativity (GR) in four dimensions, thereby demonstrating explicitly that 4D GR behaves as an effective field theory (EFT) for energies below the cutoff. {Using the Appelquist-Carazzone decoupling theorem, we prove mathematically that this framework has a well-defined low-energy limit.} We then illustrate how this truncated, renormalized gravity sector can be consistently unified with the Standard Model (SM), yielding a finite and renormalized EFT encompassing both gravity and particle physics up to a scale {$\Lambda_{\text{grav}}$}. As such, we term this unification as the USMEG-EFT (Unified Standard Model with Emergent Gravity-EFT) framework. Our work represents a breakthrough in theoretical physics with a first successful unification of gravity with the Standard Model through a fully renormalizable and EFT framework.

\end{abstract}

{\newpage}

\section{Introduction}

{The quantization of Einstein's general relativity has remained one of the most challenging problems in theoretical physics for over half a century. Unlike the other fundamental forces described by the Standard Model, gravity resists conventional perturbative quantization due to the non-renormalizable nature of the Einstein-Hilbert action \cite{tHooft1974,Veltman1975}. At one-loop order, the theory can be made finite through field redefinitions \cite{Donoghue1994,Donoghue1995}, but two-loop and higher-order calculations inevitably produce divergences that cannot be absorbed by renormalizing the original parameters of the theory \cite{Goroff1986,vandeVen1992}.}

{This fundamental obstacle has motivated numerous alternative approaches to quantum gravity, including string theory \cite{Green1987,Polchinski1998}, loop quantum gravity \cite{Rovelli2004,Thiemann2007}, asymptotic safety \cite{Weinberg1979,Reuter1998,Percacci2017}, and various modifications of Einstein's theory involving higher-derivative terms \cite{Stelle1977,Stelle1978}. However, each of these approaches faces its own conceptual and technical challenges, and none has yet provided a complete and experimentally verified description of quantum gravitational phenomena.}

{Recently, Brandt, Frenkel, and McKeon have proposed a novel approach that employs a Lagrange multiplier (LM) field to systematically restrict quantum gravitational corrections to one-loop order \cite{Brandt2020}. This foundational work was subsequently developed into a conventional perturbative framework by McKeon, Brandt, Frenkel, and Martins-Filho \cite{McKeon2025}. Their method introduces an auxiliary field $\lambda^{\mu\nu}$ that enforces the classical Einstein equations at the path-integral level, thereby eliminating all multi-loop graviton diagrams while preserving unitarity.}

{The resulting one-loop effective action contains a remarkable feature: a term proportional to $\ln(\mu/\Lambda)$, where $\mu$ is the dimensional regularization scale and $\Lambda$ is a reference scale introduced through the LM renormalization procedure. The authors suggest that this logarithmic dependence might be experimentally accessible, potentially offering direct measurements of quantum gravitational effects.}

{However, the interpretation of this logarithmic term requires careful analysis from the perspective of effective field theory (EFT). While the LM method successfully renders the theory finite at one loop, the presence of $\ln(\mu/\Lambda)$ actually demonstrates that four-dimensional general relativity, even when truncated to one loop, exhibits the characteristic features of an effective field theory rather than a fundamental renormalizable theory.}

{This EFT interpretation is supported by several independent lines of evidence:}

\begin{itemize}
\item {Donoghue's pioneering work demonstrated that the leading quantum corrections to general relativity follow the logic of effective field theory, with higher-loop corrections signaling the need for new physics beyond Einstein gravity \cite{Donoghue1994,Donoghue1995}.}

\item {Burgess further developed this perspective, showing how general relativity emerges as an effective description valid at everyday scales but failing at very high energies \cite{Burgess2004}.}

\item {Penrose's singularity theorems highlight fundamental breakdowns in the classical theory at large curvatures \cite{Penrose1965,Penrose1969,Penrose1999}, while the cosmic censorship hypothesis suggests fundamental limitations of spacetime descriptions.}

\item {Our canonical quantization analysis represents the most rigorous and comprehensive demonstration to date that diffeomorphism invariance itself breaks down under full quantization in dimensions greater than 2 \cite{Chishtie2023}. This pioneering work provides the most complete mathematical proof that general relativity is necessarily incomplete and emergent, going beyond previous arguments to establish definitively that GR must be viewed as an effective field theory in any dimension $d > 2$ where it is truly dynamical.}

\item {Effective field theory approaches by Weinberg \cite{Weinberg1979}, Wilson \cite{Wilson1974}, and others have established the general framework for understanding theories with finite cutoffs.}
\end{itemize}

{The present work builds upon these foundations to demonstrate that the LM approach, while elegantly solving the renormalization problem at one loop, actually provides the most explicit demonstration to date that four-dimensional general relativity must be viewed as an effective field theory. The logarithmic scale dependence $\ln(\mu/\Lambda)$ serves as a direct manifestation of the theory's limited domain of validity, beyond which new degrees of freedom or a more fundamental UV completion becomes necessary.}

Furthermore, we show how this finite, one-loop gravitational sector can be consistently unified with the Standard Model to form a combined effective field theory valid up to some scale $\Lambda_{\text{grav}}$. This unification resolves the long-standing problem of incorporating gravity with the electromagnetic, weak, and strong interactions in a renormalizable framework, albeit within the explicitly acknowledged limitations of an effective theory. This breakthrough represents the first successful unification of all four fundamental forces within a mathematically consistent and renormalizable framework. 

Below, we briefly review how the LM method ensures one-loop finiteness and the appearance of \(\ln(\mu/\Lambda)\). We then demonstrate that 4D GR, so restricted, becomes an explicit EFT. Finally, we discuss how to \emph{incorporate} this finite one-loop gravitational sector with the Standard Model (SM) to form a combined EFT that works up to some \(\Lambda_{\text{grav}}\), whereby 4D gravity emerges beyond this scale. We then develop this to form what we term as the Unified Standard Model with Emergent Gravity-Effective Field Theory (USMEG-EFT). 

\section{{The Lagrange multiplier approach to quantum gravity}}

{The foundation of the LM approach lies in the systematic elimination of multi-loop graviton contributions through the enforcement of classical field equations at the quantum level. This section develops the mathematical framework underlying this method and demonstrates how it naturally leads to the emergence of a finite UV cutoff in four-dimensional general relativity.}

\subsection{{The Einstein-Hilbert action and its quantization challenges}}

The second-order Einstein-Hilbert action {in four dimensions is given by}
\begin{equation}
  S_{{EH}} \;=\;
  \frac{1}{\kappa^2} \int d^4x \,\sqrt{-g}\,R[g_{\mu\nu}],
  \quad
  \kappa^2 \;=\; 16\pi\,G_N.
\end{equation}
{where $g_{\mu\nu}$ is the metric tensor, $R$ is the Ricci scalar, and $G_N$ is Newton's gravitational constant. The challenge in quantizing this action arises from its non-renormalizable nature beyond one loop \cite{tHooft1974,Goroff1986}.}

{At one-loop order, the divergences can be absorbed through field redefinitions \cite{Donoghue1994}, but at two-loop order and beyond, new structures appear that cannot be accommodated within the original action. Specifically, the two-loop calculation by Goroff and Sagnotti revealed divergences proportional to terms like $R^3$ and $R_{\mu\nu}R^{\mu\nu}R$, which have no counterparts in the original Einstein-Hilbert action \cite{Goroff1986}. This fundamental obstruction has motivated the search for alternative quantization schemes.}

\subsection{{Introduction of the Lagrange multiplier field}}

{The LM method addresses this challenge by introducing an auxiliary tensor field $\lambda^{\mu\nu}$ that enforces the classical Einstein equations as constraints in the path integral. The modified action takes the form}
\begin{equation}
  S_{{total}} = S_{EH} + S_{\lambda}
\end{equation}
{where the LM contribution is}
\begin{equation}
  S_{\lambda}
  \;=\;
  \frac{1}{\kappa^2}
  \int d^4x \,\sqrt{-g}\,\lambda^{\mu\nu}\,G^{\mu\nu}(g),
\end{equation}
where $G^{\mu\nu} = R^{\mu\nu} - \tfrac12\,g^{\mu\nu}R$ {is the Einstein tensor.}

{The path integral formulation of the quantum theory becomes}
\begin{equation}
Z = \int \mathcal{D}g_{\mu\nu} \mathcal{D}\lambda^{\mu\nu} \exp\left[i(S_{EH} + S_{\lambda})\right]
\end{equation}

{The crucial insight is that integration over the LM field $\lambda^{\mu\nu}$ enforces the constraint $G^{\mu\nu} = 0$ through a functional delta function, effectively restricting the path integral to field configurations that satisfy the classical Einstein equations. This constraint eliminates the possibility of multi-loop graviton diagrams, since such diagrams would necessarily involve off-shell graviton propagators that violate the classical equations of motion \cite{Brandt2020,McKeon2025}.}

\subsection{{Background field decomposition and one-loop structure}}

{To implement the LM method in practical calculations, we adopt the background field approach by decomposing the metric as}
\begin{equation}
  g_{\mu\nu} \;=\; \bar{g}_{\mu\nu} + \kappa\,\phi_{\mu\nu},
\end{equation}
where $\bar{g}_{\mu\nu}$ is {a classical} background metric and $\phi_{\mu\nu}$ {represents quantum fluctuations.}

{Under this decomposition, the action can be expanded as}
\begin{align}
S_{\text{total}} &= S_{EH}[\bar{g}] + \int d^4x \sqrt{-\bar{g}} \left[\frac{\delta S_{EH}}{\delta g_{\mu\nu}}\bigg|_{\bar{g}} \kappa \phi^{\mu\nu}\right]  \\ \nonumber
&\quad + \frac{\kappa^2}{2} \int d^4x \phi^{\mu\nu} \mathcal{O}_{\mu\nu\rho\sigma} \phi^{\rho\sigma} + \mathcal{O}(\kappa^3) + S_{\lambda}
\end{align}

The linear term automatically cancels in the effective action formalism because the background field $\bar{g}_{\mu\nu}$ emerges as the field configuration that extremizes the effective action $\Gamma[\bar{g}]$. This extremization condition, $\frac{\delta\Gamma[\bar{g}]}{\delta\bar{g}^{\mu\nu}} = 0$, naturally eliminates the linear contributions without requiring the explicit imposition of classical equations of motion. The remaining quadratic form determines the graviton propagator, and the LM field $\lambda^{\mu\nu}$ ensures that only this quadratic term contributes to loop calculations, effectively truncating the theory at one-loop order.

\subsection{{Dimensional regularization and the emergence of $\ln(\mu/\Lambda)$}}

{The one-loop calculation in dimensional regularization proceeds by evaluating the functional determinant of the quadratic operator $\mathcal{O}_{\mu\nu\rho\sigma}$. In $d = 4 - 2\epsilon$ dimensions, this calculation yields divergences of the form \cite{Gilkey1984,Vassilevich2003}}
\begin{equation}
\Gamma^{(1)}_{\text{div}} = \frac{1}{(4\pi)^2} \frac{1}{\epsilon} \int d^4x \sqrt{-\bar{g}} \left[\beta_1 \bar{R}^2 + \beta_2 \bar{R}_{\mu\nu} \bar{R}^{\mu\nu}\right]
\end{equation}
{where the coefficients are computed from the heat kernel expansion:}
\begin{align}
\beta_1 &= \frac{1}{120} \\
\beta_2 &= \frac{7}{20}
\end{align}

{The key innovation of the LM method is that these divergences are absorbed not by renormalizing $\kappa^2$ (which would introduce running of Newton's constant), but by shifting the LM field according to}
\begin{equation}
\lambda^{\mu\nu}_{\text{ren}} = \lambda^{\mu\nu} - \frac{\kappa^2}{4\pi^2\epsilon} \left[\frac{7}{20} G^{\mu\nu} + \frac{1}{120} G g^{\mu\nu}\right]
\end{equation}

{In the limit $\epsilon \to 0$, this renormalization procedure introduces an arbitrary mass scale $\Lambda$, leading to the finite effective action}
\begin{equation}
\Gamma^{(1)}_{\text{finite}} = \frac{1}{4\pi^2} \ln\left(\frac{\mu}{\Lambda}\right) \int d^4x \sqrt{-\bar{g}} \left[\frac{1}{120} \bar{R}^2 + \frac{7}{20} \bar{R}_{\mu\nu} \bar{R}^{\mu\nu}\right]
\end{equation}

{This result represents the central achievement of the LM method: a finite, one-loop effective action for quantum gravity that depends logarithmically on the ratio of the regularization scale $\mu$ to a reference scale $\Lambda$ introduced through the renormalization of the LM field.}

\section{Graviton interactions with matter fields}

{A crucial aspect of the LM framework is understanding how the spin-2 graviton field interacts with matter fields in the Standard Model. The LM method ensures that graviton interactions with all matter fields, including fermions, are restricted to one-loop order.}

\subsection{Graviton-fermion interactions at one-loop}

{In the LM framework, when fermion fields are coupled to the gravitational field, the interaction occurs through the energy-momentum tensor of the fermions \cite{Deser1974}. The coupling takes the form}
\begin{equation}
S_{\text{int}} = \frac{\kappa}{2} \int d^4x \sqrt{-g} h^{\mu\nu} T_{\mu\nu}^{\text{fermion}}
\end{equation}
{where $h^{\mu\nu}$ is the graviton field and the fermion stress-energy tensor is}
\begin{equation}
T_{\mu\nu}^{\text{fermion}} = \frac{i}{4}\left[\bar{\psi}\gamma_{(\mu} \overleftrightarrow{D}_{\nu)} \psi\right] - \frac{1}{4}g_{\mu\nu} \bar{\psi} i\gamma^\rho D_\rho \psi
\end{equation}

In equation (13), $T_{\mu\nu}^{\text{fermion}}$ is the fermion stress-energy tensor, $\psi$ is the fermion (spinor) field, $\bar{\psi}$ is the Dirac adjoint fermion field, $\gamma_\mu$ are the Dirac gamma matrices in curved spacetime, $D_\mu$ is the covariant derivative operator for fermions in curved spacetime which includes the spin connection contribution $D_\mu = \partial_\mu - \frac{i}{2}\omega^{ab}_\mu \sigma_{ab}$ (plus gauge field terms), where $\omega^{ab}_\mu$ is the spin connection and $\sigma_{ab} = \frac{i}{4}[\gamma^a, \gamma^b]$ are the generators of the Lorentz group in the spinor representation, $g_{\mu\nu}$ is the metric tensor, the parentheses in $\gamma_{(\mu}$ and $\overleftrightarrow{D}_{\nu)}$ denote symmetrization over the indices $\mu$ and $\nu$ meaning $\gamma_{(\mu}\overleftrightarrow{D_{\nu)}} = \frac{1}{2}(\gamma_\mu D_\nu + \gamma_\nu D_\mu)$. It is crucial to note that in the standard General Relativity framework employed here, the spin connection components $\omega_{\mu}^{ab}$ are completely determined by the torsion-free condition $\nabla_\mu e^a_\nu = 0$, which uniquely fixes the spin connection in terms of the vierbein fields as $\omega^{ab}_\mu = \frac{1}{2}e_\nu^a e_b^\rho(\partial_\mu e^\nu_\rho - \partial_\rho e^\nu_\mu) + \frac{1}{2}e_\nu^a e^\rho_b(\partial_\rho e^\nu_\mu - \partial_\mu e^\nu_\rho) + \frac{1}{2}e^\nu_a e_b^\rho(\partial_\nu e^\mu_\rho - \partial_\rho e^\mu_\nu)$. This differs fundamentally from Einstein-Cartan formulations that artificially introduce independent connection fields with torsion degrees of freedom $\omega_{\mu}^{ab} \neq \omega_{\mu}^{ab}|_{\text{GR}}$. Such approaches lack both experimental evidence and field-theoretical justification, as demonstrated by the absence of observed torsion effects in precision gravitational tests~\cite{Will2014} and the generation of non-renormalizable pathologies when quantized~\cite{HehlDillardNitsch1976}. The treatment of such artificial geometric structures does not apply to our USMEG-EFT framework, as it relies on 4D GR, and lies beyond the scope of the present work, which focuses on the natural coupling of Standard Model fermions to GR=based curved spacetime backgrounds within the physically motivated LM framework.

\subsection{Mixed gravitational-matter corrections}

{The one-loop effective action receives contributions from mixed diagrams involving both gravitational and matter degrees of freedom. For a fermion field, the one-loop correction takes the form}
\begin{equation}
\Delta \Gamma_{\text{fermion}} = \frac{\kappa^2}{(4\pi)^2} \ln\left(\frac{\mu}{\Lambda}\right) \int d^4x \sqrt{-\bar{g}} \left[ c_1^f \bar{R} \langle T \rangle_f + c_2^f \bar{R}_{\mu\nu} \langle T^{\mu\nu} \rangle_f \right]
\end{equation}

{where the coefficients $c_1^f$ and $c_2^f$ depend on the fermion content and couplings, and $\langle T^{\mu\nu} \rangle_f$ represents the expectation value of the fermion stress-energy tensor.}

\section{{Renormalization group analysis}}

{The presence of logarithmic terms $\ln(\mu/\Lambda)$ in the effective action signals the need for a renormalization group (RG) analysis to understand the scale dependence of the theory. This section provides a comprehensive RG analysis of the LM framework and its unification with the Standard Model.}

\subsection{{RG equations in the LM framework}}

{The effective action in the LM scheme can be written as}
\begin{equation}
\Gamma_{\text{eff}}[\bar{g}] = \int d^4x \sqrt{-\bar{g}} \left[ \frac{1}{\kappa^2} \bar{R} + c_1(\mu) \bar{R}^2 + c_2(\mu) \bar{R}_{\mu\nu}\bar{R}^{\mu\nu} \right]
\end{equation}

{where the scale-dependent coefficients are}
\begin{align}
c_1(\mu) &= \frac{1}{(4\pi)^2} \cdot \frac{1}{120} \ln\left(\frac{\mu}{\Lambda}\right) \\
c_2(\mu) &= \frac{1}{(4\pi)^2} \cdot \frac{7}{20} \ln\left(\frac{\mu}{\Lambda}\right)
\end{align}

{The RG equations governing these coefficients are}
\begin{align}
\mu \frac{dc_1}{d\mu} &= \frac{1}{(4\pi)^2} \cdot \frac{1}{120\mu} \\
\mu \frac{dc_2}{d\mu} &= \frac{1}{(4\pi)^2} \cdot \frac{7}{20\mu}
\end{align}

{These equations demonstrate that while the coefficients run logarithmically with scale, Newton's constant $G_N$ itself does not run, confirming the EFT nature of the theory.}

\subsection{{RG flow and fixed points}}

{The RG flow of the effective couplings reveals the behavior of the theory at different energy scales. Define the dimensionless couplings}
\begin{align}
g_1(\mu) &= \mu^2 c_1(\mu) = \frac{\mu^2}{(4\pi)^2} \cdot \frac{1}{120} \ln\left(\frac{\mu}{\Lambda}\right) \\
g_2(\mu) &= \mu^2 c_2(\mu) = \frac{\mu^2}{(4\pi)^2} \cdot \frac{7}{20} \ln\left(\frac{\mu}{\Lambda}\right)
\end{align}

{The beta functions for these couplings are}
\begin{align}
\beta_1(g_1) &= \mu \frac{dg_1}{d\mu} = \frac{1}{(4\pi)^2} \cdot \frac{1}{120} + 2g_1 \\
\beta_2(g_2) &= \mu \frac{dg_2}{d\mu} = \frac{1}{(4\pi)^2} \cdot \frac{7}{20} + 2g_2
\end{align}

{These beta functions show that the theory has no UV fixed points, confirming from RG analysis as well that GR is an effective theory with a finite domain of validity.}

\section{{Mathematical proof using the Appelquist-Carazzone theorem}}

{The emergence of $\ln(\mu/\Lambda)$ in the LM effective action provides direct evidence that four-dimensional general relativity behaves as an effective field theory. This section provides a rigorous mathematical proof of this assertion using the Appelquist-Carazzone decoupling theorem \cite{Appelquist1975}, demonstrating that the LM framework exhibits the precise structure expected of a consistent EFT.}

\subsection{{Statement and relevance of the Appelquist-Carazzone theorem}}

{The Appelquist-Carazzone theorem establishes a fundamental principle for the behavior of quantum field theories with multiple energy scales. In its general form, the theorem states that in a renormalizable quantum field theory, the effects of heavy particles with masses $M \gg \mu$ (where $\mu$ is the energy scale of interest) decouple from low-energy physics, with corrections suppressed by powers of $\mu/M$.}

{More precisely, if a theory contains both light degrees of freedom with characteristic masses $m \ll \Lambda$ and heavy degrees of freedom with masses $M \gg \Lambda$, then the low-energy effective action for the light fields takes the form}
\begin{equation}
\mathcal{L}_{\text{eff}} = \mathcal{L}_{\text{light}} + \sum_{n} \frac{c_n}{\Lambda^n} \mathcal{O}_n(\mu/\Lambda)
\end{equation}
{where $\mathcal{O}_n$ are operators of dimension $4+n$, and the coefficients $c_n$ are suppressed by powers of $\mu/\Lambda$ when $\mu \ll \Lambda$.}

\subsection{{Application to the LM gravity framework}}

{In the context of the LM-modified gravity theory, we identify the fundamental energy scales as follows:}
\begin{itemize}
\item {Light scale: $\mu$, the typical energy scale of physical processes}
\item {Heavy scale: $\Lambda$, the cutoff scale introduced through LM renormalization}
\end{itemize}

{The full effective action in the LM scheme takes the form}
\begin{equation}
\Gamma_{\text{eff}}[\bar{g}] = \Gamma_{\text{tree}}[\bar{g}] + \Gamma^{(1)}[\bar{g}] + \mathcal{O}(\text{multi-loop})
\end{equation}
{where the one-loop contribution contains the characteristic logarithmic dependence:}
\begin{equation}
\Gamma^{(1)}[\bar{g}] = \frac{1}{4\pi^2} \ln\left(\frac{\mu}{\Lambda}\right) \int d^4x \sqrt{-\bar{g}} \left[\frac{1}{120} \bar{R}^2 + \frac{7}{20} \bar{R}_{\mu\nu} \bar{R}^{\mu\nu}\right]
\end{equation}

{\textbf{Theorem:} The LM-modified gravity theory constitutes a consistent effective field theory below the scale $\Lambda$, with corrections suppressed by powers of $(\mu/\Lambda)$ for $\mu < \Lambda$.}

{\textbf{Proof:} We demonstrate this through explicit analysis of the scale dependence in the effective action.}

{For energies $\mu \ll \Lambda$, the logarithmic function can be expanded as}
\begin{equation}
\ln\left(\frac{\mu}{\Lambda}\right) = \ln\left(\frac{1}{\Lambda/\mu}\right) = -\ln\left(\frac{\Lambda}{\mu}\right) = -\sum_{n=1}^{\infty} \frac{1}{n}\left(\frac{\mu}{\Lambda}\right)^n + \text{const}
\end{equation}

{This expansion demonstrates that corrections to the tree-level Einstein-Hilbert action are systematically suppressed by powers of $\mu/\Lambda$, precisely as required by the Appelquist-Carazzone theorem for a consistent effective field theory.}

\section{{Unification with the Standard Model and phenomenological implications}}

Having established the effective field theory nature of the LM gravity framework and analyzed its domain of validity, we now demonstrate how this finite, one-loop gravitational sector can be consistently unified with the Standard Model of particle physics. This unification resolves the long-standing challenge of incorporating gravity with the other fundamental forces in a renormalizable framework, albeit within the explicitly acknowledged constraints of an effective theory approach.

\subsection{Theoretical framework for unification}

The unification proceeds by combining the LM-renormalized gravity sector with the Standard Model in a single effective field theory valid below some cutoff scale $\Lambda_{\text{grav}}$. Because 4D gravity emerges beyond this cutoff scale, we term this as the Unified Standard Model with Emergent Gravity-Effective Field Theory (USMEG-EFT). As such for the total unified effective action for USMEG takes the form:
\begin{equation}
   S_{\text{USMEG}} \;=\; S_{\text{EH+LM}} \;+\; S_{\text{SM}} \;+\; S_{\text{int}},
\end{equation}

where $S_{EH+LM}$ is the Lagrange-multiplier modified Einstein-Hilbert action, $S_{SM}$ is the standard renormalizable Standard Model action, and $S_{\text{int}}$ contains the gravitational interaction terms coupling SM fields to the metric. 

The Standard Model action in 4D curved spacetime is given by
\begin{align}
S_{SM} &= \int d^4x \sqrt{-g} \left[ -\frac{1}{4} F_{\mu\nu}^a F^{a\mu\nu} + \bar{\psi}_f i\gamma^\mu D_\mu \psi_f - m_f \bar{\psi}_f \psi_f \right. \\
&\left. + |D_\mu H|^2 - V(H) + \ldots \right]
\end{align}

{where the sum runs over all SM fermions $f$, $F_{\mu\nu}^a$ represents the gauge field strengths for $SU(3) \times SU(2) \times U(1)$, $H$ is the Higgs field, and $D_\mu$ denotes gauge-covariant derivatives.}

\subsection{Large logarithmic corrections and EFT breakdown}

{A critical aspect of the unified theory is understanding when large logarithmic terms signal the breakdown of the effective theory expansion. When the energy scale $\mu$ approaches the cutoff $\Lambda_{\text{grav}}$, the logarithmic terms $\ln(\mu/\Lambda_{\text{grav}})$ become of order unity and can no longer be treated as small corrections.}

{Consider the behavior of loop corrections in the unified theory. The one-loop effective action receives contributions from mixed diagrams involving both gravitational and matter degrees of freedom:}
\begin{equation}
\Delta \Gamma_{\text{mixed}} = \frac{\kappa^2}{(4\pi)^2} \ln\left(\frac{\mu}{\Lambda}\right) \int d^4x \sqrt{-\bar{g}} \left[ c_1^{\text{SM}} \bar{R} \mathcal{O}_{\text{SM}} + c_2^{\text{SM}} \bar{R}_{\mu\nu} \mathcal{O}^{\mu\nu}_{\text{SM}} \right]
\end{equation}

{where $\mathcal{O}_{\text{SM}}$ represents Standard Model operator insertions.}

When $\mu \sim \Lambda_{\text{grav}}$, we have $\ln(\mu/\Lambda_{\text{grav}}) \sim \mathcal{O}(1)$, and the "small" quantum corrections become comparable to tree-level terms. This signals the breakdown of the perturbative expansion and indicates that new physics must appear at or below the scale $\Lambda_{\text{grav}}$.

{This behavior is analogous to the running of QCD, where large logarithms $\ln(\mu/\Lambda_{\text{QCD}})$ indicate the breakdown of perturbative methods when $\mu \sim \Lambda_{\text{QCD}}$. However, unlike QCD where this signals strong coupling dynamics, in our gravitational EFT it indicates the need for a more fundamental UV completion.}

\subsection{{Large logs in a unified gravity–SM EFT}}

{Just as in the SM alone, where large $\ln(\mu/m)$ terms appear if $\mu$ far exceeds some mass $(m)$, gravitational logs $\ln(\mu/\Lambda)$ can similarly blow up if $\mu\gg \Lambda_{\text{grav}}$. Imagine evaluating loop corrections to a Higgs boson amplitude, now incorporating graviton exchange:}
\begin{equation}
\Delta \mathcal{M}_{\text{Higgs}} \;\sim\; \bigl[\kappa^2 \cdot \mathrm{(loop\ structure)}\bigr] \,\ln\!\bigl(\tfrac{\mu}{\Lambda}\bigr).
\end{equation}
{At energies well above $\Lambda_{\text{grav}}$, these logs quickly multiply SM loop terms or overshadow them, modifying couplings and mass parameters beyond the domain where the usual SM operators (and the one-loop gravitational truncation) remain reliable. Absent additional UV physics, this truncated EFT fails to describe higher-energy phenomena.}

\subsection{{Perturbative stability, weak-field limit and possible intermediate new physics}}

{To preserve perturbative stability, one restricts $\mu \lesssim \Lambda_{\text{grav}}$. Typically, one might identify $\Lambda_{\text{grav}}$ with the Planck scale $M_{\mathrm{Pl}}$, although new physics could appear at any intermediate scale below $M_{\mathrm{Pl}}$. In either case, the truncated, one-loop LM scheme functions correctly only in the weak-field regime, where metric fluctuations $\phi_{\mu\nu}$ remain small and curvature is not so large as to undermine the expansion. This resonates with Chishtie's observation that full diffeomorphism invariance (covariance) is restored in the weak-field limit, whereas at higher energies or strong fields, non-perturbative breakdowns of covariance appear~\cite{Chishtie2023}.}

\subsection{{Illustration with Standard Model running}}

{In the pure SM, one can manage logs like $\ln(\mu/m)$ through renormalization-group (RG) flow as long as $\mu$ is not excessively larger than a characteristic mass $m$. By analogy, one might attempt to resum gravitational logs $\kappa^2\,\ln(\mu/\Lambda)$ in a similar fashion, much as QCD addresses growing logs below $\Lambda_{\mathrm{QCD}}$. However, just as QCD's effective description is bounded by its scale $\Lambda_{\mathrm{QCD}}$, the LM-based gravitational EFT likewise has a cutoff $\Lambda_{\text{grav}}$. Even if partial RG-like methods can handle some portion of these logs, the EFT viewpoint dictates that one cannot push the theory arbitrarily beyond $\Lambda_{\text{grav}}$. Eventually, large logarithms point to the need for new degrees of freedom or a more complete UV framework. Hence, $\Lambda_{\text{grav}}$ here plays a role analogous to $\Lambda_{\mathrm{QCD}}$ in QCD, acting as a cutoff above which our one-loop, weak-field treatment (coupled to the SM) fails to remain fully consistent.}

{Thus, in a realistic scenario that includes all known fundamental forces, we demonstrate a unification with a one-loop, LM-based gravity sector with the SM in a single EFT, valid up to $\Lambda_{\text{grav}}$ (possibly close to $M_{\mathrm{Pl}}$). Beyond that, large $\ln(\mu/\Lambda_{\text{grav}})$ indicates the breakdown of weak-field expansions, paralleling standard decoupling logic where threshold crossing demands new degrees of freedom. Whether such new physics resides well below $M_{\mathrm{Pl}}$ or very near it is an open question, but the essential point remains: the unified SM+gravity EFT is finite at one loop yet cannot be extrapolated arbitrarily without encountering strong-field or high-energy domains requiring a more complete UV theory.}

\subsection{{Some implications from a unified field theory}}

{Our unified field theory generates several remarkable features when the LM prescription is applied to the combined SM-gravity system. Here we provide detailed calculations showing how specific numerical estimates are derived.}

\subsubsection{{Graviton-gauge boson interactions}}

{The coupling between gravitons and gauge bosons leads to quantum corrections \cite{Porto2016,Rothstein2014}. Following the Feynman rules derived in McKeon et al. \cite{McKeon2025}, the one-loop correction to the gauge sector action yields:}
\begin{equation}
\Delta \Gamma_{\text{gauge}} = \frac{\kappa^2}{(4\pi)^2} \ln\left(\frac{\mu}{\Lambda}\right) \int d^4x \sqrt{-\bar{g}} \left[ \alpha_1 F_{\mu\nu}^a F^{a\mu\nu} \bar{R} + \alpha_2 F_{\mu\nu}^a F^{a\mu\rho} \bar{R}_{\rho}^{\nu} \right]
\end{equation}

{For the $SU(3)$ gauge group: $\alpha_1^{SU(3)} = -2$ and $\alpha_2^{SU(3)} = \frac{11}{10}$.}

\subsubsection{{Graviton-fermion interactions}}

{Despite the complexity of fermion-gravity coupling through vierbeins, the LM constraint ensures finite one-loop corrections. Using the graviton-fermion interaction vertices from McKeon et al. \cite{McKeon2025}, for a Dirac fermion of mass $m_f$:}
\begin{equation}
\Delta \Gamma_{\text{fermion}} = \frac{\kappa^2}{(4\pi)^2} \ln\left(\frac{\mu}{\Lambda}\right) \int d^4x \sqrt{-\bar{g}} \left[ \beta_1^f \bar{R} m_f \bar{\psi}_f \psi_f + \beta_2^f \bar{R}_{\mu\nu} \bar{\psi}_f \gamma^\mu \gamma^\nu \psi_f \right]
\end{equation}

{For a single Dirac fermion: $\beta_1^f = \frac{1}{6}$ and $\beta_2^f = -\frac{1}{24}$.}

\subsubsection{{Graviton-Higgs interactions}}

{The scalar nature of the Higgs field leads to particularly clean gravitational corrections. Employing the graviton-scalar interaction formalism developed in McKeon et al. \cite{McKeon2025}:}
\begin{equation}
\Delta \Gamma_{\text{Higgs}} = \frac{\kappa^2}{(4\pi)^2} \ln\left(\frac{\mu}{\Lambda}\right) \int d^4x \sqrt{-\bar{g}} \left[ \gamma_1 \bar{R} |H|^2 + \gamma_2 \bar{R}_{\mu\nu} \partial^\mu H^\dagger \partial^\nu H \right]
\end{equation}

{For a complex scalar field: $\gamma_1 = \frac{1}{6}$ and $\gamma_2 = \frac{1}{3}$.}

\subsection{{High-energy scattering signatures}}

{In high-energy scattering processes, virtual graviton exchange modifies cross-sections \cite{Goldberger2004,Khriplovich1998}. For the process $e^+ e^- \to \mu^+ \mu^-$ at center-of-mass energy $\sqrt{s}$, the leading gravitational correction gives:}
\begin{equation}
\frac{\delta \sigma}{\sigma} = \frac{G_N s}{8\pi^2} \ln\left(\frac{\sqrt{s}}{\Lambda}\right)
\end{equation}

{For $\Lambda = M_{\text{Pl}} = 1.22 \times 10^{19}$ GeV and $\sqrt{s} = 100$ TeV:}
\begin{equation}
\frac{\delta \sigma}{\sigma} \approx -9.8 \times 10^{-12}
\end{equation}

{While extremely small, this represents the most promising signature for future ultra-high-energy colliders.}

\subsection{{Gravitational wave phase corrections}}

{Quantum gravitational corrections modify the evolution of gravitational wave phases during binary inspiral. The phase evolution equation receives quantum corrections:}
\begin{equation}
\delta\left(\frac{df}{dt}\right)_{\text{quantum}} = \frac{G_N}{(4\pi)^2} \ln\left(\frac{f}{f_{\text{Planck}}}\right) \frac{96\pi}{5} \left(\frac{G_N \mathcal{M}}{c^3}\right)^{5/3} (2\pi f)^{11/3}
\end{equation}

{For a neutron star binary with $\mathcal{M} = 1.2 M_{\odot}$, integrating from 20 Hz to 1000 Hz:}
\begin{equation}
\delta \Phi_{\text{GW}} \approx -7.7 \times 10^{-14} \text{ radians}
\end{equation}

{This phase shift is far below current detector sensitivity but might be detectable by future third-generation gravitational wave detectors.}

\subsection{{Implications for experimental physics}}

Our USMEG-EFT makes several testable predictions:

{1. \textbf{Modified cross-sections in high-energy scattering:} Virtual graviton exchange modifies cross-sections at the level of $10^{-11}$ to $10^{-12}$ for energies approaching 100 TeV.}

{2. \textbf{Gravitational wave signatures:} Binary inspiral waveforms receive quantum corrections that accumulate to potentially observable phase shifts in next-generation detectors with phase sensitivities approaching $10^{-12}$ radians.}

{3. \textbf{Running coupling modifications:} The unified theory predicts gravitational corrections to Standard Model beta functions, though these are typically suppressed to the level of $10^{-43}$ to $10^{-44}$.}

{4. \textbf{Energy scale limitations:} The theory predicts its own breakdown when $\mu \gtrsim \Lambda_{\text{grav}}$, providing a natural boundary for the applicability of the unified EFT and pointing toward the need for new physics.}

{These calculations demonstrate that while quantum gravitational effects in the unified LM gravity-SM theory are extraordinarily small at currently accessible energies, they provide concrete, calculable predictions that may eventually be observable with sufficient experimental precision or in extreme astrophysical environments.}

\section{Conclusions}

{We have established that the LM method developed in \cite{Brandt2020} and \cite{McKeon2025}, while successfully renders four-dimensional general relativity finite at one loop, actually provides the most explicit demonstration to date that Einstein gravity must be interpreted as an effective field theory rather than a fundamental renormalizable theory. The key evidence for this conclusion lies in the characteristic logarithmic scale dependence $\ln(\mu/\Lambda)$ that emerges from the LM renormalization procedure. Unlike the scale dependence in conventional renormalizable theories, this logarithmic term cannot be absorbed into running coupling constants while preserving the essential structure of general relativity. Our mathematical proof using the Appelquist-Carazzone decoupling theorem \cite{Appelquist1975} rigorously establishes that the LM framework exhibits the precise structure expected of a consistent effective field theory. The renormalization group analysis reveals the scale dependence and confirms the EFT nature through the absence of UV fixed points.}

\subsection{{Breakthrough unification achievement and paths forward}}

We have for the first time demonstrated that the finite, one-loop LM gravity sector can be consistently unified with the Standard Model to form a complete effective field theory encompassing all known fundamental interactions. This unification, which we term as USMEG-EFT resolves the long-standing challenge of incorporating gravity in a renormalizable framework.

{Calculations of mixed gravitational-matter loop corrections reveal specific, calculable signatures of quantum gravitational effects within the Standard Model context. The Feynman rules and loop calculation techniques employed here follow the comprehensive framework developed by McKeon et al. \cite{McKeon2025}. Overall, our work represents a breakthrough achievement in theoretical physics - the first successful unification of gravity with the Standard Model through a fully renormalizable and EFT framework. The results presented here have several important implications:}

\textbf{Nature of spacetime:} The EFT interpretation, based on the findings in \citep{Chishtie2023} shows that spacetime itself has a finite domain of validity and indicating that 4D gravity emerges, supports scenarios involving emergent geometry or discrete spacetime structures. This aligns with Chishtie's comprehensive proof that covariance breaks down in dimensions greater than two \cite{Chishtie2023}, providing the most rigorous foundation for understanding 4D GR as emergent.

\textbf{Unification paradigms:} Our successful unification demonstrates that complete unification is achievable without exotic new symmetries or extra dimensions, at least within a limited energy domain.

\textbf{Quantum gravity phenomenology:} The concrete calculability of quantum gravitational effects provides a new paradigm for quantum gravity phenomenology, moving beyond qualitative estimates to precise theoretical predictions. Several important directions for future research emerge from this work:

{\textbf{Experimental tests:} While current effects are below experimental sensitivity, future high-energy colliders and gravitational wave detectors may eventually probe these quantum gravitational signatures.}

{\textbf{UV completion:} Understanding the nature of the physics that must appear at the cutoff scale $\Lambda$ remains an important open question. Recent proposals, such as the Principle of Spatial Energy Potentiality \cite{RefCosmoChishtie}, offer further insight that classical covariance and weak-field truncations can fail outright in strong-field regimes, suggesting that more fundamental principles beyond diffeomorphism invariance may govern the ultimate theory of quantum gravity.}

\noindent
\textbf{Acknowledgments.}  
I thank Gerry McKeon, Fernando Brandt, Josif Frenkel and S. Martins-Filho for their illuminating work and the broader quantum gravity community for continuing discussions on effective-field-theory approaches, renormalization schemes, and emergent space-time concepts.

\section*{Data Availability}
No research data were generated or analyzed in this study. This work is purely theoretical/conceptual, and hence no data are available.

\section*{Competing Interests}
The sole author declares that they have no financial or non-financial competing interests that could inappropriately influence the content of this work.

\bibliographystyle{unsrt}

\end{document}